\documentclass[aps, prL,twocolumn,amsmath,amssymb, superscriptaddress]{revtex4}
\usepackage{graphicx}
\usepackage{latexsym}
\usepackage{txfonts}
\usepackage{amsmath,amssymb} 
\usepackage{graphicx}
\usepackage{amsmath}
\usepackage{amssymb}
\usepackage{amsfonts}
\usepackage{upgreek}
\usepackage{txfonts}
\usepackage{color}
\usepackage{comment}
\usepackage{rotating}
\usepackage{mathdots}
\begin{document}

\title{
Quantum wetting transitions in two dimensions: \\
an alternative path to non-universal interfacial singularities
}

\author{Pawel Jakubczyk} 
\author{Marek Napi\'{o}rkowski} 
\affiliation{Institute of Theoretical Physics, Faculty of Physics, University of Warsaw, Pasteura 5, 02-093 Warsaw, Poland} 

\author{Federico Benitez}
\affiliation{Max Planck Institute for Solid State Research, Heisenbergstrasse 1, 70560 Stuttgart, Germany}

\date{\today}

\begin{abstract}
We consider two-dimensional ($d=2$) systems with short-ranged microscopic interactions, where interface unbinding (wetting) transitions occur in the limit of vanishing temperature $T$. For $T=0$ the transition is characterized by non-universal critical properties analogous to those established for thermal wetting transitions in $d=3$, albeit with a redefined capillary parameter $\tilde{\omega}$. 
Within a functional renormalization-group treatment of an effective interfacial model, we compute the finite temperature phase diagram, exhibiting a line of interface unbinding transitions, terminating at $T=0$ with an interfacial quantum critical point. At finite $T$ we identify distinct scaling regimes, reflecting the interplay between quantum and thermal interfacial fluctuations. A crossover line marking the onset of the quantum critical regime is described by the $d=3$ interfacial correlation-length exponent $\nu_{||}$. This opens a new way to investigate the non-universal character of $\nu_{||}$ without penetrating the true critical regime. On the other hand, the emergent interfacial quantum critical regime shows no signatures of non-universality.
\end{abstract}

\maketitle

%\section{Introduction}
Interfacial phase transitions received an enormous interest over the last decades \cite{Dietrich88, Schick90, Forgacs91, Bonn09, Parry09}. Of peculiar status are the wetting (or interface unbinding) transitions in systems characterized by short-ranged microscopic interactions. The upper critical dimension for these transitions turns out to coincide with the physical dimensionality $d=3$, yielding complex and unusual critical behavior. The critical exponents are predicted to be non-universal and depend on a single dimensionless capillary parameter $\tilde{\omega}$ \cite{Brezin83, Fisher85}. The nature of short-ranged critical wetting in $d=3$ has long been of intense dispute due to discrepancies between renormalization-group studies of effective capillary-wave Hamiltonians and  Monte-Carlo simulations of Ising-type models \cite{Binder86, Binder88, Binder89}. The latter never confirmed the prediction of non-universality, and, instead, yielded results compatible with mean-field theory (MFT). A resolution of this long-standing paradox has recently been offered by invoking non-local effects \cite{Parry04, Parry06, Parry07, Parry08, Bernardino09}, which explain the elusive nature of critical wetting by the fact that the asymptotic critical region is extremely hard to reach.   

In this paper we explore an alternative route to detect the signatures of non-universality, invoking the $T\to 0$ limit of the interfacial unbinding transition in $d=2$. Strictly at $T=0$ we find the interfacial quantum phase transition to be of very similar nature to the classical $d=3$ transition, however with a redefined capillary parameter $\tilde{\omega}$.  Our computations at $T>0$ reveal the existence of an interfacial quantum critical regime (QCR) separating the mean-field (MFTR) and quantum (QR) scaling regimes from the true critical thermal regime (TR) dominated by classical fluctuations, see Fig.~1. The shape of the crossover line marking the onset of the QCR region is described by the exponent $\nu_{||}$, which (up to a redefinition of the capillary parameter $\tilde{\omega}$) also governs the divergence of the interfacial correlation length $\xi_{||}$ at classical critical wetting in $d=3$. This correspondence potentially opens a possibility of detecting the onset of non-universality without penetrating the asymptotic critical region, for example in simulations of the $d=2$ quantum Ising model in a transverse field, subject to a boundary field.     
\begin{figure}
 \includegraphics[width=9.0cm]{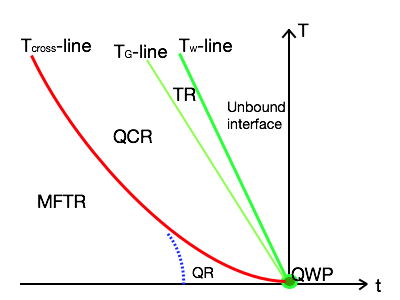}
\caption{(color online) Phase diagram illustrating the scaling regimes in the vicinity of the quantum interface unbinding transition in $d=2$. The line of thermal phase transitions ($T_w$-line) belongs to the $d=2$ Ising universality class and terminates at $T=0$ with the quantum wetting point (QWP). The thermal regime (TR) squashed between the $T_w$-line and the Ginzburg line ($T_G$-line) is dominated by classical fluctuations. In the quantum critical regime (QCR) both thermal and quantum fluctuations contribute to the leading behavior given by Eq.~(\ref{xi_QCR}). The shape of the crossover line ($T_{cross}$-line) $T_{cross}\sim |t|^{\nu_{||}}$ is governed by the critical exponent ($\nu_{||}$) analogous to the one describing the divergence of the interfacial correlation length $\xi_{||}$ at $d=3$ classical critical wetting. The interfacial critical singularities at $T=0$ are the same as for classical critical wetting in $d=3$, albeit with a different capillary parameter $\tilde{\omega}^q$ given by Eq.~(\ref{omegaq}). The dotted blue line separates the critical $T=0$ regime (QR) from the mean-field regime (MFTR). Its precise shape at $T>0$ was not computed here.  }
%\end{center}
\end{figure}  

Consider a quantum system in an equilibrium thermodynamic state in the vicinity of coexistence between two bulk phases ($\alpha$ and $\gamma$). Phase $\alpha$ is infinitesimally stable. The system is enclosed by a rigid, plain wall. The microscopic interactions are such that phase $\gamma$ is preferentially adsorbed at the boundary, so that an interface separates the layer of the $\gamma$ phase condensed at the wall, and the bulk-stable phase $\alpha$. The system is sufficiently far from any bulk critical point, so that fluctuations in the bulk phases are small. The separation $l$ of the $\alpha-\gamma$ interface from the wall can be controlled by varying temperature $T$ as well as a non-thermal parameter $t$ (for example the wall properties, see below). For particular values of these parameters the interface unbinds ($l\to\infty$), which is accompanied by the divergence of interfacial fluctuations (the capillary waves).   
By virtue of the quantum-classical mapping, scaling properties of  bulk quantum systems at $T=0$ are related to those characterizing classical systems, but in elevated dimensionality. Following, vicinity of a bulk quantum phase transition \cite{Sachdev_book} is described by the Landau-Ginzburg type action: 
\begin{equation}
S_{LGW}^b[\phi]=\int_0^{\beta \hbar} d\tilde{t}\int d^dr \left[\frac{K_{\tilde{t}}}{2} \left(\partial_{\tilde{t}}\phi\right)^2 + \frac{K_{\mathbf{r}}}{2} \left(\partial_{\mathbf{r}}\phi\right)^2 + \frac{m^2}{2!}\phi^2 + \frac{\lambda}{4!}\phi^4   \right]\;, 
\label{LGW_bulk}
\end{equation} 
where we focus on transitions involving only discrete symmetry breaking (single-component, scalar bulk order parameter field $\phi$). The system is therefore equivalent to a $d+1$-dimensional anisotropic classical system, where non-zero temperature corresponds to a finite-size effect in the ''time'' ($\tilde{t}$) direction. The stiffness parameters $K_{\tilde{t}}$ and $K_{\mathbf{r}}$ as well as the mass $m$ and the interaction coupling $\lambda$ may be related to parameters of an underlying truly microscopic model such as the quantum Ising model in a transverse field \cite{Sachdev_book}.

Once the system is in contact with the wall, translational symmetry becomes broken, which is accounted for by an additional term 
\begin{equation}
S_{LGW}^s[\phi] = \int_0^{\beta \hbar} d\tilde{t}\int d^{d}r \left[ -h_1\phi-\frac{1}{2}g\phi^2\right]\delta (r_d) \;. 
\label{LGW_surface}
\end{equation} 
The field expansion in Eq.~(\ref{LGW_surface}) is truncated at quadratic order, keeping the term involving the surface field $h_1$ and the surface enhancement parameter $g$ (see e.g. \cite{Binder83}). Additionally, presence of the wall restricts the integration along one of the spatial directions ($r_d$) in Eq.~(\ref{LGW_bulk}) to the interval $[0^-,\infty]$. 
The theory defined by 
\begin{equation}
S[\phi]=S_{LGW}^b[\phi]+S_{LGW}^s[\phi] 
\label{LGW}
\end{equation} 
 is now subject to the coarse-graining procedure proposed by Jin and Fisher \cite{Jin93, Rejmer96}.  
 This amounts to an approximate, mean-field-like tracing over the bulk order-parameter field $\phi(\tilde{t}, {\bf r}=({\bf x},r_d))$ under the constraint of fixed interface position at a distance $l(\tilde{t}, \mathbf{x})$ above the wall. The mean-field procedure is justified by the fact, that the bulk degrees of freedom are massive, and the modes responsible for the critical interfacial singularities, captured by the field $l({\tilde{t}, \bf x})$, and are not integrated out here. The relation between the resulting effective interfacial action $\mathcal{S}_{eff}[l]$ and the order-parameter action (\ref{LGW}) is given by 
 \begin{equation}
 e^{-\mathcal{S}_{eff}[l]/\hbar} = \int^{l}\mathcal{D}\phi e^{-S[\phi]/\hbar} \;,
\end{equation}
where the symbol $\int^{l}$ indicates that the functional integral is restricted to the configurations of $\phi(\tilde{t},\mathbf{r})$ satisfying the condition of pre-specified field $l(\tilde{t},\mathbf{x})$. The Fisher-Jin type coarse-graining yields the quantum variant of the capillary-wave model
\begin{equation}
\mathcal{S}_{eff}[l({\bf x}, \tilde{t})] = \int_0^{\beta\hbar}d\tilde{t}\int d^{d-1}x\left[V(l)+\frac{\sigma}{2}\left(\partial_{{\bf x}}l\right)^2+\frac{Z}{2}\left(\partial_{{\tilde t}}l\right)^2 \right] \;.
\label{Hamiltonian}
\end{equation}
The effective interactions between the interface and the wall are described by the local effective potential $V(l)$. For the presently relevant case of short-ranged microscopic interactions, $V(l)$ can be expanded in powers of $e^{-l/\xi_b}$, with $\xi_b$ being a bulk correlation length, which may also be considered as a short-distance (real space) cutoff of the theory. The coefficient $t$ of the leading term in this expansion may be used to tune the transition and is related to the surface field $h_1$ (see Eq.~(\ref{LGW_surface})). It acts as a non-themal control parameter of our system. The parameter $\sigma$ is the usual classical interfacial stiffness coefficient, while the quantity $Z$ controls quantum fluctuations. The partition function is now given by 
$ 
\mathcal{Z}=\int \mathcal{D}le^{-\mathcal{S}_{eff}[l]/\hbar}\;. 
$
The classical limit of the theory is recovered by restricting $l$ to be uniform in imaginary time $\tilde{t}$. The model of Eq.~(\ref{Hamiltonian}) was employed in Ref.~\cite{Jakubczyk12} to study quantum interface unbinding transitions in $d\geq 3$, where the quantum ($T \to 0$) limit of the theory is mean-field like. Here we focus on the completely different case $d=2$, where the limit $T\to 0$ is nontrivial.
  
We emphasise that the quantum variant of the capillary-wave model can be established on grounds as firm as its classical limit. We note that the procedure of Ref.~\cite{Jin93} leads to an $l$-dependent stiffness coefficient $\sigma(l)$, which destabilises the renormalization-group flow \cite{Jin293, Boulter98} towards a first-order transition. The non-local theory \cite{Parry04, Parry06, Parry07, Parry08, Bernardino09} sheds light on the nature of these problems, and, in fact, implies that disregarding the $l$-dependent corrections to $\sigma$ yields a much better local approximation of the more complete theory. We also mention a significant limitation of the present theory: The Fisher-Jin derivation of the capillary-wave model assumes that there are no bulk soft modes present. This excludes, {\it inter allia}, all bulk phases characterized by breaking of a continuous symmetry and therefore hosting Goldstone excitations. This limitation applies as much to both quantum and classical systems. Also note that the interface unbinding transitions are most often studied in the context of fluids which are described within the classical framework. Experimental accessibility of fluid-type systems fully justifies such state of the art. On the other hand, the Cahn argument \cite{Cahn77} for the existence of wetting-type transitions holds also for low-temperature systems requiring quantum treatment. Obvious examples are (quantum) magnets and superconductors.

It is possible to deduce some of the important features of the phase diagram of the system defined by Eq.~(\ref{Hamiltonian}) (see Fig.~1) by generalizing the classical reasoning of Lipowsky and Fisher \cite{Lipowsky87}. A heuristic estimate of the difference $\Delta f$ between the free energy of the free and bound interface involves three terms: the energy change due to the interaction with the wall $V(\bar l)$, a contribution due to thermal fluctuations $U_{fl}^{cl}(\bar l)$ and an analogous quantum term $U_{fl}^{q}(\bar l)$ . The quantity $U_{fl}^{cl}$ accounts for an energetic contribution due to elasticity and an entropic repulsion from the wall. In $d=2$ one finds  $U_{fl}^{cl}\sim (\beta\xi_{||})^{-1}$. For the quantum contribution we obtain $U_{fl}^{q}\sim \hbar\sqrt{\sigma/Z}\xi_{||}^{-2}$, where we dropped a logarithmic correction.   
The free energy difference 
\begin{equation}
\label{Delta_f}
\Delta f = V + U_{fl}^{cl} + U_{fl}^q
\end{equation} 
may now be inspected in different asymptotic regimes. First compare the relative importance of the fluctuation contributions. The quantum fluctuations remain negligible as long as $U_{fl}^{cl} \gg U_{fl}^q$, which is equivalent to $k_BT\gg\hbar\xi_{||}^{-1}=\hbar|t|^{\nu_{||}}$. In consequence, the condition $k_BT=\hbar|t|^{\nu_{||}}$ defines the crossover line $T_{cross}$ in Fig.~1. Next: in the limit of vanishing temperatures $\beta\to \infty$ we neglect $U_{fl}^{cl}$ in Eq.~(\ref{Delta_f}). Mean-field theory yields $V\sim |t|^2$ and $\nu_{||}=1$. The dependence on $t$ then drops out, which reflects the marginal character of the theory at $T=0$. Determination of the shape of the Ginzburg line at asymptotically small $T$ (the blue dotted line in Fig.~1) requires going beyond the present level of analysis and is not achieved here. At higher $T$ or asymptotically close to the thermal transition line we drop the term $U_{fl}^q$ in Eq.~(\ref{Delta_f}). The condition $V \approx U_{fl}^{cl}$ then yields the thermal Ginzburg line $T_G$. We obtain $T_G(t)\sim |t|$. Note that the $T_G$-line persists down to $T=0$, since $U_{fl}^{cl} \gg U_{fl}^q$ at any finite $T$ sufficiently close to the transition. This heuristic picture gives a hint on the origin of the different scaling regimes in Fig.~1 and the shape of the $T_{cross}$ and $T_G$ lines. 

We now discuss a functional renormalization group calculation, which confirms this picture and yields the behavior of interfacial properites in the scaling regimes QCR and QR in Fig.~1. The TR regime is governed by the well known classical $d=2$ wetting singularities. We apply the functional RG theory formulation of Ref.~\cite{Wetterich93}, adapted to the interfacial transitions in Ref.~\cite{Jakubczyk11}. The point of departure is the exact flow equation for the momentum-scale ($k$) dependent effective action $\Gamma_k[l]$, i.e. the  generating functional for the one-particle-irreducible vertex functions \cite{Wetterich93}:
\begin{equation}
\partial_k \Gamma_k[l] = \frac{1}{2}{\rm Tr}\frac{\partial_k R_k({\bf q}, \omega)}{\Gamma_k^{(2)}[l]+R_k({\bf q}, \omega)}\;. 
\label{Wetterich_eq}
\end{equation}
The quantity $\Gamma_k[l]$ may be understood as a free energy including fluctuation modes with momentum above the scale $k$. In Eq.~(\ref{Wetterich_eq}) $R_k({\bf q}, \omega)$ is the momentum ($\bf q$) and/or frequency ($\omega$) cutoff function, added to the inverse propagator to regularize the theory in the infrared (${\bf q}, \omega \to 0$), and $\Gamma_k^{(2)}[l]$ denotes the second functional derivative of $\Gamma_k[l]$. The trace sums over momenta and Matsubara frequencies $\omega_n=2\pi n/(\beta\hbar)$ (Fourier-conjugate to the imaginary time $\tilde{t}$): 
$
{\rm Tr} = \frac{1}{\beta\hbar}\sum_n \int\frac{d^{d-1}q}{(2\pi)^{d-1}}\;. 
\label{Tr}
$ 
By varying the momentum cutoff scale $k$ below the short-distance limit ($k=\Lambda\sim \xi_b^{-1})$, fluctuations of decreasing momentum are included and in the limit $k\to 0$ the quantity $\beta^{-1}\Gamma_{k\to 0}[l]$ coincides with the full Gibbs free energy. On the other hand, $\Gamma_{k=\Lambda}[l] = \mathcal{S}_{eff}[l]/\hbar$. The flow equation (\ref{Wetterich_eq}) does not admit an analytical solution. Ref.~\cite{Jakubczyk11} discusses its relation to other formulations of the RG theory for interfacial phenomena. The approximation we now apply amounts to imposing a parametrization of $\Gamma_k[l]$ by the following form:
\begin{equation}
\Gamma_k[l]=\frac{1}{\hbar}\int_0^{\beta\hbar}d\tilde{t}\int d^{d-1}x\left[V_k(l)+\frac{\sigma}{2}\left(\partial_{{\bf x}}l\right)^2+\frac{Z}{2}\left(\partial_{{\tilde t}}l\right)^2 \right]\;. 
\label{ansatz}
\end{equation}
This implies locality of the scale-dependent effective potential $V_k(l)$ and neglects renormalization of the stiffness coefficients $\sigma$ and $Z$. Since the anomalous dimension $\eta=0$ for interface unbinding transitions \cite{Lipowsky87}, $\sigma$ and $Z$ are expected to acquire only finite renormalization in an exact theory, which has no impact on the critical singularities. We also disregard higher-order gradient terms generated in course of the flow. The present approach recovers all the results of classical local capillary-wave RG theory \cite{Fisher85, Lipowsky87}, both in the linear and non-linear regime, for a discussion see \cite{Jakubczyk11}. By plugging the ansatz (\ref{ansatz}) into Eq.~(\ref{Wetterich_eq}), we obtain the flow equation for the effective potential
\begin{equation}
\partial_kV_k(l)/\hbar=\frac{1}{2}{\rm Tr}\frac{\partial_k R_k({\bf q}, \omega)}{\left[\sigma {\bf q}^2 +Z \omega^2+V_k''(l)\right]/\hbar +R_k({\bf q}, \omega)}\;,
\label{LPA_equation}
\end{equation}
which is the starting point for our analysis, focusing on $d=2$. We also specify the cutoff 
\begin{equation}
R_k({\bf Q}) = \frac{\sigma}{\hbar} \left(k^2-{\bf Q}^2\right)\theta \left(k^2-{\bf Q}^2\right) 
\label{Litim_cutoff}
\end{equation}
where ${\bf Q}=(\sqrt{\frac{Z}{\sigma}}\omega, {\bf q})$. None of the conclusions of the paper are sensitive to this choice.

We first analyze the RG flows at $T=0$, where $\frac{1}{\beta\hbar}\sum_n \to \int\frac{d\omega}{2\pi}$.
The substitution $\omega=\sqrt{\frac{\sigma}{Z}}q_0$ brings the flow equation (\ref{LPA_equation}) to an isotropic form. 
%\begin{equation}
%\partial_kV_k(l)/\hbar=\frac{1}{2}\sqrt{\frac{\sigma}{Z}}\int\frac{d^d Q}{(2\pi)^d}\frac{\partial_k R_k({\bf Q})}{\left[\sigma {\bf Q}^2+V_k''(l)\right]/\hbar+R_k({\bf Q})} 
%\label{LPA_T=0}
%\end{equation}
%where ${\bf Q}=(q_0,{\bf q})$. We now specify the cutoff function $R_k({\bf Q})$ \cite{Litim01}, 
%\begin{equation}
%R_k({\bf Q}) = \frac{\sigma}{\hbar} \left(k^2-{\bf Q}^2\right)\theta \left(k^2-{\bf Q}^2\right) 
%\label{Litim_cutoff}
%\end{equation}
%which facilitates straightforward evaluation of the integrals. We emphasise that, when specified to $d=2$, the results of this section do not depend on the particular form of $R_k({\bf Q})$. This may be proven along the line developed in Ref.~\cite{Jakubczyk11} for the classical case. The integration in Eq.~(\ref{LPA_T=0}) yields 
%\begin{equation} 
%\partial_kV_k(l)/\hbar=\sqrt{\frac{\sigma}{Z}}\frac{S^{d-1}}{d(2\pi)^d}\frac{\sigma k^{d+1}}{\sigma k^2 +V_k''(l)}\;, 
%\label{LPA_T=0_integrated}
%\end{equation}
%where $S^{d-1}$ is the surface area of a $d-1$-dimensional sphere. We now specify to $d=2$ and observe that, up to factors, the flow equation (\ref{LPA_T=0_integrated}) is equivalent to the one studied in the context of classical capillary-wave theory in Ref.~\cite{Jakubczyk11} in the case $d=3$. 
Subsequently the flow equation is linearized in $V''(l)$, recovering a diffusion-type flow equation 
\begin{equation} 
\partial_k V_k(l)=\alpha (k) - \frac{\hbar}{4\pi\sqrt{Z\sigma}k}V_k''(l) \;. 
\label{linearized_flow}
\end{equation}
The term $\alpha (k)$ merely shifts the flowing effective potential without modifying its shape, and may be dropped. Importantly, the factor multiplying $V_k''(l)$ does not depend on the form of $R_k({\bf q},\omega)$ for the present case $d=2$. By the standard rescaling one now recovers the classical linear RG theory of wetting transitions in $d=3$ \cite{Fisher85}, where the dimensionless classical capillary parameter 
$
\tilde{\omega}=\frac{k_B T}{4\pi\sigma\xi_b^2} 
$
 governing the non-universal critical behavior becomes replaced by 
\begin{equation} 
\tilde{\omega}^q = \frac{\hbar}{4\pi\sqrt{Z\sigma}\xi_b^2} \;. 
\label{omegaq}
\end{equation}
The critical behavior for interface unbinding in $d=2$ and $T=0$ is therefore non-universal and can be related to the one predicted for $d=3$ in the case of transitions driven by thermal fluctuations. This also determines the asymptotic properties of the system in the QR regime, where thermal fluctuations are effectively inactive (see below). 

The classical limit of our model is recovered by restricting the path integral to configurations uniform in $\tilde{t}$, or (equivalently) considering only the thermal ($\omega=0$) contribution to the trace in Eq.~(\ref{LPA_equation}). 
Executing the momentum integrals in Eq.~(\ref{LPA_equation}) and performing the rescaling of variables: 
$z = Dlk^{\frac{3-d}{2}},\, v_k(z) = C V_k(l)k^{1-d}  $
 brings the flow equation to the scale-invariant form. We choose the following rescaling factors
 $C=\frac{\beta(2\pi)^{d-1}(d-1)(3-d)}{2S^{d-2}},\, D^2=C\sigma$. 
  The flow equation then takes the following form:
 \begin{equation}
 \partial_s v_k(z)=(d-1)v_k(z)+\frac{3-d}{2}zv_k'(z)+\frac{d-3}{2}\frac{1}{1+ v_k''(z)}\;.
 \label{rescaled_classical_flow}
 \end{equation}
 Above we introduced $s=-\log k/\Lambda$. Note that the specific form of the last term of Eq.~(\ref{rescaled_classical_flow}) does depend on the particular choice of the cutoff (\ref{Litim_cutoff}). The standard formulation \cite{Lipowsky87} of the non-linear functional renormalization group theory of classical interface unbinding transitions yields a term $\sim\ln(1+v_k''(z))$ instead of $\frac{1}{1+ v_k''(z)}$ in the flow equation Eq.~(\ref{rescaled_classical_flow}). The equation studied in Refs~\cite{Lipowsky87} is recovered in our formulation upon making the sharp cutoff choice $R_k(q)=\lim_{\gamma\to\infty}\sigma k^2\gamma/\hbar \theta(k^2-q^2)$ instead of Eq.~(\ref{Litim_cutoff}). The fixed-point equation $\partial_s v_k(z)=\partial_s v_{FP}(z)=0$ involves a single dimensionless parameter $g=\frac{2(d-1)}{3-d}$ and can be solved analytically in the asymptotic regimes. At $z\to 0^+$ one finds $v_{FP}(z)=c^< z^{-g}+\dots $, while for $z\to\infty$ the fixed-point equation may be linearised and finds solution in terms of the Hermite and hypergeometric functions. Numerical integration of the fixed-point equation yields a line of fixed point potentials parametrized by $c^<$, revealing a landscape of possible interfacial critical phenomena in $d<3$ (not restricted to short-ranged interactions). The fixed-points relevant to wetting with short-ranged forces are identified by the Gaussian tails at large $z$.
 The picture emergent in $d=2$ turns to be equivalent to the one obtained by Lipowsky in the context of amphiphilic membranes \cite{Lipowsky88} in $d=3$.  Apart from different coefficients [stemming from the kinetic term $\sim (\partial_{\bf x}^2 l)^2 $ relevant for membranes - compare Eq.~(\ref{Hamiltonian})], the present fixed-point equation differs from the one of Ref.~\cite{Lipowsky88} in the particular form of the nonlinear term. As shown by an explicit numerical integration of our fixed-point equation, this does not change the content of the analysis in any way. We remark that the present analysis of the classical sector is relevant to the thermal regime (TR) in Fig.~1 and refer to the work by Lipowsky~\cite{Lipowsky88} for a complete characteristic of the fixed-point potentials and a discussion of their meaning. Also note that this regime is well explored by exact studies of the Ising model in $d=2$ \cite{Abraham80}.   
 
We now investigate the system for low, but finite $T$ and off the truly critical thermal regime. By performing the integration in Eq.~(\ref{LPA_equation}) we obtain the flow equation 
 \begin{equation}
 \partial_kV_k(l)=\beta_k^{cl}(l)+ \beta_k^{q}(l) 
 \label{flow_eq}
 \end{equation}
 involving the classical ($\omega=0$) term
 \begin{equation}
 \beta_k^{cl}(l)=A_d\beta^{-1}\frac{\sigma k^d}{\sigma k^2+V_k''(l)}
 \end{equation} 
 and the quantum contribution ($\omega\neq 0$) 
 \begin{equation}
 \beta_k^{q}(l)=2A_d\beta^{-1}\frac{\sigma k}{\sigma k^2+V_k''(l)}\sum_{n=1}^{n_{max}}\left[k^2-\frac{Z}{\sigma}\left(\frac{2\pi n}{\beta \hbar}\right)^2\right]^{\frac{d-1}{2}}\;.
 \end{equation}
  Above we introduced $A_d= \frac{S^{d-2}}{(2\pi)^{d-1}(d-1)}$ ($S^d$ being the surface area of the $d$-dimensional unit sphere) and $n_{max}\in \mathbb{N}$ defined so that the expression in the square bracket is positive (exclusively) for $n=1,\dots, n_{max}$. Rescaling according to the classical dimensions yields 
  \begin{align}
   \partial_sv=  (d-1)v+  \frac{3-d}{2}zv'+ \;\;\;\;\;\;\;\;\;\; \;\;\;\;\;\;\;\;\;\; \;\;\;\;\;\;\;\;\;\;\;\;\;\;\;\;\;\;\;\;\;\;\;\\
   \frac{1}{1+v''}\frac{d-3}{2}\left[1+2\sum_{n=1}^{n_{max}}\left(1-\tilde T^2n^2\right)^{\frac{d-1}{2}}\right]\;, \nonumber 
   \label{Flow}
  \end{align}
 where 
 $
 \tilde{T} = \sqrt{\frac{Z}{\sigma}}\frac{2\pi}{\beta\hbar}k^{-1}
 $
 may be given the interpretation of a flowing temperature \cite{Millis93, Jakubczyk08}. We also abbreviated notation by dropping the arguments of the flowing rescaled potential. Obviously, $\tilde{T}$ grows monotonously in course of the flow for any $T>0$. In the following analytical solution we use the strategy adapted from the theory of bulk quantum criticality \cite{Millis93, Bauer11}: The flow starts at small $T$ and the right-hand-side of the flow equation (\ref{flow_eq}) is approximated by its $T\to 0$ limit for scales $k$ such that $\tilde{T}<1$. The condition $\tilde{T}=1$ marks the crossover scale $k_{cross}$, below which the most significant contribution to the flow originates from the classical term. This is signalled by $n_{max}$ decreasing below 1 and vanishing of the quantum term $\beta_k^{q}(l)$. We terminate the flow once reaching the perturbative scale $k_p$, where the (rescaled) correlation length becomes of the microscopic order set by $\Lambda^{-1}$. One may then use the mean-field calculation and subsequently scale back to the original lengthscales. The condition $k_p=k_{cross}$ marks the crossover line in the ($t -T$) phase diagram. Indeed if $k_p>k_{cross}$ the flow never reaches the classical scaling regime, and the universal physics can be considered to be the same as at $T=0$. On the other hand, if $k_p<k_{cross}$, the flow traverses both the classical and quantum sectors and the computed physical quantities are influenced by both quantum and thermal fluctuations.    

Following the procedure outlined above, we use the $T=0$ form of the flow equation for $\tilde{T}<1$. We integrate the linearized flow Eq.~(\ref{linearized_flow}) until reaching either the perturbative scale $k_p$ where we terminate the flow, or the crossover scale $k_{cross}$, where we switch to the classical scaling. The initial stage of the analysis largely paralellizes Ref.~\cite{Fisher85}. The bare effective potential is taken as 
$
V_\Lambda (\tilde{z})=A_\Lambda (\tilde{z})+R_\Lambda(\tilde{z})+W_\Lambda(\tilde{z})
$    
and involves the attractive tail  $A_\Lambda (\tilde{z})=t e^{-\tilde{z}}\theta(\tilde{z})$, the repulsive tail $R_\Lambda (\tilde{z})=b e^{-2\tilde{z}}\theta(\tilde{z})$, and the short-distance soft wall repulsion $W_\Lambda(\tilde{z})=w\theta(-\tilde{z})$. Here $\tilde{z}=l/\xi_b$. The linearized flow can be integrated via Fourier transform. 
 Our analysis indicates that, as long as $k_p>k_{cross}$, the system features the singularities characteristic for $d=3$ classical critical wetting, with the replacement $\tilde{\omega}\rightarrow\tilde{\omega}^q$. The condition $k_p=k_{cross}$ determines the crossover line in the $(t-T)$ phase diagram. We find  
 \begin{equation}
 T_{cross}\approx | t|^{\nu_{||}(\tilde{\omega}^q)}\;,
 \end{equation} 
 where $\nu_{||}(x)=1/(1-x)$ for $x<1/2$ and $\nu_{||}(x)=1/(\sqrt{2}-\sqrt{x})^2$ for $x\in (1/2,2)$. The exponent $\nu_{||}$ diverges for $\tilde{\omega}^q\to 2^-$, where the region in the $(t-T)$ plane characterized by the $T=0$~-type behavior becomes exponentially suppressed.
 This calculation confirms and complements our previous phenomenological reasoning.  
% We observe that, in fact, the exponent describing the shape of the crossover line $T_{cross}(\tau)$ can be related to the correlation length exponent for classical wetting in $d=3$ with the replaced capillary parameter $\tilde{\omega}^{cl}\to \tilde{\omega}^{q}$. %This is the central result of this subsection. Namely, for $\tilde{\omega}^q<2$ we find 
 %\begin{equation}
 %T_{cross}\approx |\tau|^{\nu_{||}(\tilde{\omega}^q)}\;,
 %\end{equation} 
 %where $\nu_{||}(x)=1/(1-x)$ for $x<1/2$ and $\nu_{||}(x)=1/(\sqrt{2}-\sqrt{x})^2$. The exponent $\nu_{||}$ diverges for $\tilde{\omega}^q\to 2^-$, where the region in the $(T-\tau)$ plane characterised by the $T=0$-type behavior becomes exponentially suppressed. 
 Also observe the crucial difference of these findings as compared to the case of bulk quantum criticality, where the crossover line is fully determined by dimensionality $d$ and the dynamical exponent $z$.   

We finally explore the case $k_p<k_{cross}$, where the flow encounters the crossover scale before reaching the perturbative regime. This resolves the asymptotic behavior in the QCR regime in Fig.~1. Down to the scale $k_{cross}$ the flow is integrated as above. The resultant effective potential $v_{k_{cross}}(z)$ then serves as an initial condition for the second stage of the flow spanning between $k_p$ and $k_{cross}$. This regime is dominated by the thermal contribution and we here retain only the classical term in Eq.~(18). Upon linearization, the resulting flow equation is integrated down to $k_p$. Our calculation gives  
\begin{equation}
k_p\approx \frac{G_1}{\log[G_2 t^2\beta^{2(1-\tilde{\omega}^q)}]} 
\label{k_p}
\end{equation}
with $G_1 = k_BT/(\pi\sigma\xi_b^2)$ and $G_2=(2\sigma b)^{-1}[Z^{1/2}\sigma^{-1/2}2\pi\xi_b/\hbar]^{2\tilde{\omega}^q}$. This yields 
\begin{equation}
\xi_{||}\sim k_p^{-1} \sim \frac{\log |t|}{T} + \frac{\log(1/T)}{T}. 
\label{xi_QCR}
\end{equation} 
This behavior is remarkably different from the one found in all the other regimes depicted in Fig.~1. Upon picking any fixed $T>0$ and scanning the dependence of $\xi_{||}$ on $t$, the onset of the QCR may be detected by the fall of the exponent $\nu_{||}$ to 0 upon crossing the $T_{cross}$-line. This constitutes a clear and new prediction, which is  also independent of our ability to resolve the critical regions in the QR and TR regions in Fig.~1 in a realistic situation. 

The above calculation uses a linearized form of the RG flow equation. This is justified by the fact that we remain separated from the true critical regime governed by the soft thermal fluctuations. The well-developed theory of bulk quantum critical phenomena described by $\phi^4$-type theories indicates that the analogous bulk quantum critical regime is correctly captured by a low-order calculation (essentially retaining only the mass renormalization via the tadpole diagram) missing the Wilson-Fisher fixed point.   
 
As summarized in Fig.~1, our results point at the existence of a number of scaling regimes in the interfacial phase diagram of a generic two-dimensional system featuring an interface unbinding transition at $T\to 0$. Two aspects of the system exhibit non-universal behavior cognate to the celebrated classical critical wetting in $d=3$, namely the QR regime and the shape of the crossover line $T_{cross}(t)$. The QCR regime is described by a very weak dependence of the correlation length $\xi_{||}$ on $t$ which results in a sudden decrease of the exponent $\nu_{||}$ upon crossing the $T_{cross}$-line or the $T_G$ line along any isotherm. The critical singularities are stronger and the phase diagram significantly richer as compared to the previously studied almost trivial case $d\geq 3$. We believe the QCR regime should be detectable in simulations of the quantum Ising model subject to a boundary field. It is also tempting to speculate about the possibility of diagnosing the non-universal character of the $\nu_{||}$ exponent by inspecting the shape of the crossover line. The proposed avenue seems superior to the previous attempts in that it does not require a resolution of the critical region. On the other hand, the present study does not give a definite answer as to the temperatures required to observe the quantum scaling.       
 
 \emph{Acknowledgement} We acknowledge funding by the Polish Ministry of Science and Higher Education via grant IP2012 014572.

% The quantum transitions are however characterised by a different capillary parameter involving $\hbar$ and the time-stiffness coefficient $Z$. This incomplete fulfilment of the quantum-classical mapping can in fact be expected from the outset, since the thermal %capillary parameter vanishes identically in the quantum limit $T\to 0$. On the other hand, once we consider an arbitrarily small, finite temperature, the quantum fluctuations become gapped and the critical behaviour is supposed to be dominated by the classical %fluctuations sufficiently close to the transition. It is now interesting to investigate the interplay between the quantum and thermal effects at finite, but arbitrarily low $T$ (see Sec.~VI)

\end{document}